\documentclass[manuscript]{aastex}

\slugcomment{Accepted for publication in The Astrophysical Journal}

\shorttitle{3-14 Micron Spectroscopy of Comets}
\shortauthors{Sitko et al.}

\begin{document}

\title{3 - 14 Micron Spectroscopy of Comets C/2002 O4 (H\"{o}nig),  \\
    C/2002 V1 (NEAT), C/2002 X5 (Kudo-Fujikawa), \\
    C/2002 Y1 (Juels-Holvorcem), 69P/Taylor, \\
    and the Relationships among Grain Temperature, \\ Silicate Band Strength
    and Structure among Comet Families}

\author{Michael L. Sitko\altaffilmark{1}}
\affil{Department of Physics, University of Cincinnati, Cincinnati OH 45221}
\email{sitko@physics.uc.edu}

\author{David K. Lynch\altaffilmark{1} and Ray W. Russell\altaffilmark{1}}
\affil{The Aerospace Corporation, Los Angeles, CA 90009}
\email{david.k.lynch@aero.org, ray.w.russell@aero.org}

\and

\author{Martha S. Hanner}
\affil{Department of Astronomy, University of Massachusetts, Amherst, MA 01003}
\email{mhanner@astro.umass.edu}

\altaffiltext{1}{Visiting Astronomer, NASA Infrared Telescope Facility, operated by the University of Hawaii under contract with the National Aeronautics and Space Administration.}

\begin{abstract}
We report 3 - 13 $\mu$m spectroscopy of 4 comets observed between August 2002 and February 2003: C/2002 O4 (H\"{o}nig) on August 1, 2002, C/2002 V1 (NEAT) on Jan. 9 and 10, 2003, C/2002 X5 (Kudo-Fujikawa) on Jan. 9 and 10, 2003, and C/2002 Y1 (Juels-Holvorcem) on Feb. 20, 2003. In addition, we include data obtained much earlier on 69P/Taylor (February 9, 1998) but not previously published. For Comets Taylor, H\"{o}nig, NEAT, and Kudo-Fujikawa, the silicate emission band was detected, being approximately 23\%, 12\%, 15\%, and 10\%, respectively, above the continuum. The data for Comet Juels-Holvorcem were of insufficient quality to detect the presence of a silicate band of comparable strength to the other three objects, and we place an upper limit of 24\% on this feature. The silicate features in both NEAT and Kudo-Fujikawa contained structure indicating the presence of crystalline material. The shape of the silicate feature at a projected distance of 1900 km from the nucleus of Kudo-Fujukawa was nearly identical to that centered on the nucleus, indicating that the grain size population had not been measurable modified by the time it had reached that distance. Combining these data with those of other comets, we confirm the correlation between silicate band strength and grain temperature of Gehrz \& Ney (1992) and Williams et al. (1997) for dynamically new and long period comets, but the majority of Jupiter family objects may deviate from this relation. Despite the weakness of the silicate band in Kudo-Fujikawa, its structure resembles the bands seen in dynamically new and long period objects with substantially stronger features. The limited data available on Jupiter family objects suggest that they may have silicate bands that are slightly different from the former objects. Finally, when compared to the silicate emission bands observed in pre-main sequence stars, the dynamically new and long period comets most closely resemble the more evolved stellar systems, while the limited data (in quantity and quality) on Jupiter family objects seem to suggest that these have spectra more like the less-evolved stars. Higher quality data on a larger number of Jupiter family objects are needed to confirm (or reject) this trend.
\end{abstract}


\section{Introduction}

One of the major goals of astronomy today is to understand the origin and evolution of solar systems, including our own. In the same way that stellar astrophysics is tied to our detailed understanding of the Sun for guidance, our ability to understand the evolution of extrasolar planetary systems in anything more than a general and tentative way will require detailed knowledge of how the most easily accessed system has developed. One of the key pieces of information that is required to model the formation and evolution of the solar system is the Òboundary conditionÓ of its initial state. And while understanding the specific details of the initial collapse of the protosolar cloud may remain an unreachable goal at the present time, much of the physical and chemical nature of our solar system during its first $10^{6}$ years, when it was only 0.02\% of its present age, remains in the form of solids stored beyond the orbit of Neptune where the equilibrium temperature drops below 50 K. These materials, true interstellar material that survived intact, material that condensed out of the early nebular gas, and material processed during the earliest stages of nebular development, are accessible in the form of comets.

The presence of silicate material in comets was first confirmed by \citet{maas70}, who unambiguously detected the emission band of silicates in the spectrum of Comet C/1969 Y1 Bennett using filter photometry with a spectral resolution R = $\lambda$/$\Delta\lambda\sim 10$. Since then, numerous bright comets have been observed using broad-band and intermediate-band photometry \citep{ney74a,ney74b,gn92,att86a,att86b,msh87}, and this technique continues to be used to this day. While it lacks the mineralogical diagnostic power obtainable with higher spectral resolution, its broad wavelength coverage (2-23 $\mu$m and often extending to 0.7 $\mu$m) has made it a powerful tool for understanding the nature of cometary dust grains. Two important discoveries made with this technique are that the strength of the silicate band varies from object to object, and that most bright comets have grains that radiate at temperatures in excess of that of a blackbody at the expected equilibrium temperature for their heliocentric distance. Both effects are expected for a grain population that includes a significant fraction of the grains having sizes smaller that the wavelength of light being radiated, in this case from 3-20 $\mu$m, and the two have been found to be correlated in the emission from bright comets \citep{gn92,wil97}.

The data relevant to the population of grain sizes may be dependent on the position within the coma where the observations are made, and therefore also dependent on the beam size used in the measurement. The grain size distribution measured in the coma of 1P/Halley by the Giotto and Vega spacecraft varied noticeably as a function of distance from the nucleus and between pre- and post-encounter tracks \citep{mcd87,mcd91}. This is not surprising, as the entrainment of dust by subliming volatiles will tend to sort grains by mass (or the ratio of mass to surface area for highly porous grains), and fragmentation of grains during outflow will tend to produce small grains from large ones. Ground-based photometry of C/1973 E1 Kohoutek indicated that the grains in the anti-tail were considerably larger than those in the central coma and tail \citep{ney74a}. Such behavior is expected in comets due to the sorting by radiation pressure (see, for example, \citet{gj90}). 

To properly characterize the grain sizes, thermal properties, and mineralogical composition of relic solar system grains, spectroscopic data of a large number of comets are required. Objects ranging from short-period comets to dynamically new ones need to be included. Comets that have entered the inner solar system for the first time will have surfaces that have been exposed for billions of years to processing by cosmic rays, and may not necessarily be shedding the more pristine materials to be found deeper in their interiors. Short-period objects that have endured thousands of years (or more) of surface activity will have had this outer layer removed. However, because smaller grains are more easily entrained by sublimating volatiles from the nucleus than larger ones, their population of observable grain sizes will tend to be skewed toward larger values than what they were born with. Hence, data on a large number of objects at various stages of their evolution will be required to disentangle the various factors affecting their grain properties. In this paper we report our observations of 5 comets, and analyze the data for these and 14 others for which data at comparable spectral resolution and wavelength coverage exist.

\section{Observations}
These comets were observed over the course of four separate observing runs on the NASA Infrared Telescope Facility (IRTF) on Mauna Kea. Observations were obtained using the Aerospace Corporation's Broadband Array Spectrograph system (BASS). This instrument uses a cold beamsplitter to separate the light into two separate wavelength regimes. The short wavelength beam includes light from 2.9-6 $\mu$m, while the long-wavelength beam covers 6-13.5 $\mu$m. Each beam is dispersed onto a 58-element Blocked Impurity Band (BIB) linear array, thus allowing for simultaneous coverage of the spectrum from 2.9-13.5 $\mu$m. The spectral resolution $R = \lambda$/$\Delta\lambda$ is wavelength-dependent, ranging from about 30 to 125 over each of the two wavelength regions. At the IRTF, the circular entrance aperture of the instrument subtends 3.4 arcsec on the sky. All of the observations are calibrated relative to a set of spectral standards ($\alpha$  CMa, $\alpha$ Lyr, $\alpha$ Tau, $\alpha$ Boo, and $\beta$ Peg)  rigorously tested for internal consistency.

All of the comets were observed with the nucleus centered in the circular aperture. The details of each observation are listed in Table 1.

On 10 Jan. 2003 between 0244-0325 (UT) we also observed Kudo-Fujikawa  offset 2.6 arcsec west of the nucleus,  during which time the airmasss increased from 2.34 to 3.38, with a mean of 2.81. For these higher-airmass observations, we constructed a mean spectrum for $\alpha$ CMa by combining two observations, one at 3.20 airmass and the other at 1.53 airmass obtained later in the evening. By weighting the former by a factor of three to one over the latter, a spectrum expected for an airmass of 2.78 was obtained. 

\section{Results}

\subsection{Comet Kudo-Fujikawa}

The BASS observations of Comet Kudo-Fujikawa are shown in Figure 1a. The observed flux was the same for both nights for $\lambda < 8 \mu$m, but was systematically brighter on 9 January at longer wavelengths. Also shown in Figure 1a is an estimate of the underlying continuum, determined by fitting a gray-body planckian curve normalized to two spectral regions: 7.5-8.5 $\mu$m and 12-13 $\mu$m. For 9 January, the best fit was obtained using a continuum temperature $T_{C} = 340$ K. For 10 January, $T_{C} = 355$ K and $T_{C} = 360$ K gave adequate fits. With a heliocentric distance of 0.67 and 0.63 AU for 9 and 10 January, respectively, the observed temperatures were nearly identical to predicted equilibrium blackbody temperatures of $T_{BB} = 340$ K (9 January) and $T_{BB} = 350$ K  (10 January). 

The difference in flux between the two nights, shown in Figure 2, shows no evidence of a silicate band, and is greatest outside the wavelength range of that feature. The net change in flux is thus dominated by the change in the continuum, and must be the result of differences in the population of larger grains between the two nights. Because the comet became hotter as it faded between January 9 and 10, the change in net flux must have been the result of a diminishing production rate of larger grains between the two nights, as the increased temperature on the 10th would have increased the net flux had the mass of dust been constant. 

In order to gain some measure of the grain sizes present, we have measured the strength of the silicate band above the surrounding continuum. To obtain an unbiased measurement of the (in-band flux)/(continuum flux) ratio, we determined the mean value of the total flux between 10.0 and 11.0 $\mu$m (7 data points), where the signal is cleanest, and the shape of typical cometary silicate emission features is relatively flat. This was then divided by the underlying 8-13 $\mu$m graybody flux, assumed to represent the continuum outside the silicate emission band. The ratio is shown in Figure 3a for both values of $T_{C}$. On 9 January, the band/continuum flux ratio was 1.09 $\pm$ 0.01. The error quoted is the standard deviation of the signal in this restricted wavelength region, and includes both the random statistical errors associated with measuring the flux, as well as any real systematic variations due to band structure. For the following day, the ratio was essentially identical. Using $T = 355$ K for the underlying continuum of the January 10 observation yielded a ratio of 1.11 $\pm$ 0.01, while assuming $T = 360$ K gave a ratio of 1.09 $\pm$ 0.01. Thus, in addition to the random error of 0.01, we estimate a possible additional systematic error of 0.02 due to the uncertainty in fitting the underlying continuum. Nevertheless, it would seem that for these nights, a mean value of 1.10 $\pm$ 0.02 is reasonable. The relative weakness of the silicate emission band, coupled with grain temperatures close to that of blackbodies in equilibrium, suggests that the population of silicate grains less than 10 $\mu$m in size was small. 

The break in slope of the data at 11.2 $\mu$m in Figure 3a indicates the presence of crystalline olivine, which has been observed in numerous comets (c.f. \citet{hlr94}), but is difficult to detect in objects with a relatively weak silicate band strength. In this data, a secondary feature at 10.5 $\mu$m is also visible. This latter feature was present in Hale-Bopp \citep{dhw99} and Levy \citep{dkl92a}, and is attributed to crystalline enstatite \citep{chiha02}. When visible in comet spectra, it is usually much weaker than the olivine peak at 11.2 $\mu$m, however. Interestingly, some glass-rich chondritic interplanetary dust particles (IDPs) of presumed cometary origin show a similar feature. One such IDP (U219C11) studied by \citet{bhg92}  possessed broad bands at 10.6 and 11.2 $\mu$m of comparable strength.

On 10 January we also obtained data offset by 2.6 arcsec directly west of the nucleus, placing it on an angle 97$^{\circ}$ from the Sun-comet line. The time of the observations of the coma ranged from 0244-0322 UT, compared to 0120-0244 UT for the nucleus, a mean difference of 1.2 hours. Figure 1b shows the flux of the coma at this location and the spectrum centered on the nucleus. The fact that the spectrum obtained on the coma at this location was indistinguishable from the data on the nucleus (other than a constant multiplicative factor) indicates the grain populations had the same temperature and composition throughout the span of time it took the grains to drift the 2.6 arcsec from the nucleus Ð a projected distance of about 1900 km at $\Delta$=1.0 AU. If the dust velocity were constant and similar to that observed in Comet Hale-Bopp (0.67$ \pm$ 0.07 km s$^{-1}$; \citet{braun97}), then that projected distance corresponds to dust ejected 0.8 hours earlier. For a radial density distribution dropping off as the inverse-square of the distance (as would occur for a constant-velocity flow), the density-weighted mean distance is twice this value. Thus, to a first approximation we were sampling the same grains in both cases. If any grain fragmentation occurred during this time interval, it failed to produce any measurable change in the grain size population. However, due to the weakness of the band strength in this comet, small changes would escape detection.

\subsection{Comet NEAT}
The data on Comet NEAT obtained on 9 and 10 January 2003 are shown in Figures 1c and 3b. Fitting a graybody curve to the continuum at 7.5-8.5 $\mu$m and 12-13 $\mu$m, we obtained a mean grain temperature of $T_{C} = 290$ K for the data obtained on 9 January. On January 10, the derived temperature was $T_{C} = 280$ K, identical to that obtained by \citet{hon04}  using the COMICS spectrograph on the Subaru telescope the same night. For comparison, we also show the blackbody curve for $T_{C} = 290$ K for the January 10 data. These temperatures are approximately 30 K higher than that expected for a blackbody at the two heliocentric distance of the comet at the time of observation (1.197 AU and 1.174 UA, respectively). 

On both nights, the plateau of the silicate feature between 10 - 11 $\mu$m was in excess of 10\% above the continuum, and was bounded at the long wavelength side by the 11.2 $\mu$m peak attributed to crystalline olivine. Honda et al. used a 4-component plus blackbody model to fit their higher spectral resolution COMICS data. In their model, 9\% of the mass of the material emitting above the continuum level was in the form of crystalline olivine.

\subsection{Comet Juels-Holvorcem}
This faint (narrow-band magnitude at 10.2 $\mu$m of $3.5 \pm 0.1$) object was observed for 33 minutes, with observations being terminated by clouds and morning twilight. The spectrum and underlying continuum, with $T_{C} = 290$ K, is shown in Figure 1d. The band/continuum flux ratio, shown in Figure 3c, is very noisy. For this object, we attempted to determine the strength of the silicate feature using three different techniques. An unweighted mean of the 7 points included in the 10 - 11 $\mu$m wavelength range was $1.02 \pm 0.08$. Using a mean of the fluxes weighted by their respective errors yielded $1.03\pm 0.02$. A straight median of the data yielded 1.02. Using the most conservative value, we place an upper limit on the strength of the band of 1.25. The temperature of the underlying continuum was 38 K above the expected blackbody value for a heliocentric distance of $r = 1.219$ AU.

\subsection{Comet H\"{o}nig}
Comet H\"{o}nig was observed on August 1, 2002. As pointed out by \citet{sek03}, the behavior of this comet was quite strange, having dropped precipitously in visual brightness prior to perihelion. He has suggested that the comet was undergoing a major outburst near the time of its discovery and later suffered bulk disintegration through erosion of the nucleus. In this scenario, our observations were obtained at the time when the comet was making the transition from the phase of major dust production to a more quiescent rate, and about a week or two prior to the start of nuclear disintegration.

Comet H\"{o}nig was observed over a mere 33 minute period of time, and the data are of lower quality than that of the previous two objects. The observed flux, and the spectrum of a blackbody with $T=280$ K normalized in the same was as was done previously, is shown in Figure 1e. The resulting flux/continuum ratio, seen in Figure 3d, had a median value between 10 - 11 $\mu$m of $1.12 \pm 0.08$. The temperature of the underlying continuum was 42 K higher than that expected ($T = 237$) for a blackbody in equilibrium at the heliocentric distance of the comet ($r = 1.367$ AU) on that date.

\subsection{Comet Taylor}
The spectrum of 69P/Taylor, obtained on Feb. 9, 1998, is shown in Figure 1f. The underlying continuum had a temperature of approximately $230 \pm 20$ K, about 40 K above the expected blackbody equilibrium temperature. Using this continuum, a weak silicate feature is discernible, and the resulting band/continuum ratio is shown in Figure 3e.

\section{Grain Size and Excess Temperature}

Grains radiating at temperatures in excess of that of a blackbody in thermal equilibrium, often called ÒsuperheatÓ, is a natural consequence of grains possessing a mean emissivity in the thermal infrared that is smaller than the mean emissivity in the visible (weighted by the solar spectrum), where the radiation is absorbed. To first order the ratio of the observed color temperature of the continuum $T_{C}$ to that of a blackbody $T_{BB}$ is

$(T_{C}/T_{BB}) \approx (Q_{VIS}/Q_{IR})^{1/4}$
  
\noindent where $Q_{VIS}$ and $Q_{IR}$ are the absorption efficiency factors averaged over the incident solar spectrum and the re-emitted thermal emission, respectively (see for example \citet{dkl89}). By definition, a true blackbody has $Q_{\lambda} =1$ for all wavelengths $\lambda$, while $Q_{\lambda} =$ constant $< 1$ for a ``graybody''. The observed temperature excess over that of a blackbody could indicate either a preponderance of grains smaller than the wavelengths that they are radiating at (about 10 $\mu$m), or grains whose mean absorption coefficient (the emissivity for a radiating body) is higher in the visible (i.e. lower albedo) than in the IR due to compositional effects. For example, Hanner et al. (1997; see their Figure 1) have shown that glassy carbon spheres radiate near the equilibrium temperature if their sizes are larger than a few microns, but are significantly warmer if smaller than a micron. Similarly, increasing the absorption coefficient of the silicate grains in the 3-8 $\mu$m spectral region (increasing Fe/Mg for the silicates is but one way to do this) will tend to shift the temperature of the grains upward.

In Table 1 we list the excess temperature, defined in the same manner as \citet{gn92}) and \citet{wil97} as 
  
$S=(T_{C}/T_{BB})$

Here, it is important to remember that there can, in principle, be a significant difference in the temperature defined by the spectral shape, its color temperature, and the actual physical temperature of the grains. This is because the former is determined by the product of the Planck function and $Q_{IR}$. These quantities are identical, however, for a blackbody or ÒgraybodyÓ, where $Q_{IR}$ = constant. This is the reason we have used the spectrum of the underlying continuum to determine $T_{C}$, and for internal consistency used the same two spectral regions to define it for all objects in our sample.

Using Mie theory as a rough approximation of the absorption properties of the grains, it can be shown that in the limit where the wavelength of the light is much smaller than the size of the grain, the efficiency for absorption $Q_{abs}$ approaches unity (see, for example \citet{bh83}). If the wavelength is much larger than the grain size, then $Q_{abs}\propto 1/\lambda$. Thus for a given grain size, $Q$ will generally decrease when going from visible to infrared wavelengths, for small grains that have some absorption at visible wavelengths. The wavelength where one crosses between these two regimes is determined by the ratio
 
$\Omega = 4 \pi ka/\lambda$

\noindent where $a$ is the radius of the (spherical) grain in the same units as the wavelength, and $k$ is the imaginary component of the complex index of refraction of the grain at that wavelength \citep{lm00}. $\Omega = 1$ marks the boundary where the transition from $Q \sim 1$ to $Q < 1$ occurs, and is dependent on both the actual size distribution of the grains and their optical constants.

Because of this, we would expect that a collection of compact cometary grains composed entirely of large grains to have $Q \sim 1$ everywhere, and radiate like blackbodies, while those with a significant fraction of small grains to radiate at higher temperatures, i.e. have $S > 1$. (The situation for fluffy grains is more complicated; see \citet{xh97} and \citet{gh90} for example). Likewise, for large grains, as $Q$ becomes independent of wavelength the contrast of the silicate band to the continuum diminishes (although it is never totally lost; see \citet{min03}). This occurs when $a \sim \lambda$ for typical silicate grains with measured optical constants (Henning et al. 1995; \citet{lm00}). Hence, we would expect that the band/continuum ratio and $S$ would be correlated. 

\citet{gn92} demonstrated that such a correlation existed, based on data obtained on 9 bright comets: 1P/Halley, C/1969 Y1 Bennett, C/1973 E1 Kohoutek, C/1975 V1 West, C/1975 N1 Kobayashi-Berger-Milon, C/1980 Y1 Bradfield, C/1984 N1 Austin, C/1989 X1 Austin, and 23P/Brorsen-Metcalf. \citet{wil97}) extended this to comet C/1995 O1 Hale-Bopp. In these studies, the silicate excess expressed in magnitudes 
 
$\Delta m = 2.5log(F_{band}/F_{cont})$

\noindent is rather tightly correlated with $S$ (note that Gehrz and Ney actually plot the percent excess temperature, defined as $[S-1] x 100$).

We have calculated $\Delta$m and $S$ for the 5 comets in the present study and tabulated them in Table 1. We have also used the spectrophotometric data on a larger sample of objects and calculated these same quantities using the exact same criteria, for the comets in the following studies: 

\citet{hlr94}: C/1973 E1 Kohoutek, C/1983 H1 IRAS-Araki-Alcock, 1P/Halley,
C/1986 P1 Wilson, C/1987 P1 Bradfield, 23P/Brorson-Metcalf, C/1989 Q1
Okazaki-Levy-Rudenko, C/1989 X1 Austin, and C/1990 K1 Levy

\citet{msh96}: 19P/Borrelly, 4P/Faye, 24P/Schaumasse

\citet{dkl00}: 55P/Tempel-Tuttle

\citet{dkl02}: C/1999 T1 McNaught-Hartley

We list these quantities in Table 1, along with a classification of the comets based on their orbital parameters (DN - dynamically new, YLP - young long period, OLP - old long period, HF - Halley family, and JP - Jupiter family), using the criteria of \citet{ah95}. In a few cases, the band/continuum ratio and $T_{C}$ differ slightly from those originally reported. For example, in \citet{hlr94}, $T_{C}$ was calculated for Kohoutek using data at mid-infrared wavelengths obtained the day before the spectrophotometric data shown. The spectrophotometric data actually fall somewhat above this continuum (see their Fig. 1a). There is also some overlap with the objects in Gehrz \& Ney and Williams et al. However, Gehrz \& Ney utilized a different set of data: broad- and intermediate-resolution filter photometry. Williams also included higher-resolution spectral data for Hale-Bopp. In contrast, we have attempted to derive both the value of $T_{C}$ and the band strength using the same criteria and wavelength regions for our entire sample. The resulting relationship between $\Delta$m and $S$ is shown in Figure 4. Also shown in the figure is an error-weighted  2-dimensional fit to the data, excluding the Jupiter family objects and the peculiar Comet Wilson (see below). Including the JF comets has little effect (due to their relatively large errors) while including Comet Wilson would reduce the slope slightly. For simplicity, for those 5 objects with multiple sets of data in our sample, we have represented their location in the figure using the mean and  its standard deviation of the 2 data sets. In the case of one object (Levy) this has the effect of increasing the size of its errors, but does not alter the fit  to the data, as both fall almost precisely on the line shown. Also note that because $\Delta$m and $S$ are not entirely physically independent (the continuum level can be affected by $S$), this would tend to mathematically produce an anti-correlation and reduce the slope observed. 

In this figure, and despite the slight bias expected against it mentioned above, most of the dynamically new and long period comets exhibit the tight correlation between $\Delta$m and $S$ reported by Gehrz \& Ney and Williams et al. However, many objects in the sample are scattered to the right of this relation. This is true as a class for the JF comets in this study (Borrelly, Faye, Schaumasse, and Taylor), none of which are represented in the plots by Gehrz \& Ney or Williams et al. They deviate in band strength $\Delta$m from the fit shown in Figure 4 by 2.2 $\sigma$ (Schaumasse) to 7.6 $\sigma$ (Faye), where $\sigma$ is the standard deviation represented by the error bars in the figure. In terms of excess temperature $S$ the range is 1.4 $\sigma$ (Taylor) to 2.3 $\sigma$ (Faye). In both cases, the deviation is systematically in the same direction. IRAS Low Resolution Spectrograph observations of two other JF comets, Tempel 1 and Tempel 2 (not shown here) also exhibit excess temperature but no measurable silicate feature \citep{dkl95}, and would also deviate from the correlation defined by the DN comets in the same manner as the other JF comets in our sample. Two independent observations of the JF comet 103P/Hartley 2 using different instruments aboard the Infrared Space Observatory (ISO), separated by only one day, give conflicting results.  The observations by Colangeli et al. (1999), using ISOPHOT-S, show no silicate band, while Crovisier et al. (1999,2000), using spectra extracted from images obtained using the circular variable filter on ISOCAM, detected a small but significant emission band. So we consider this deviation to be typical of the class, and not in disagreement with the findings of Gehrz \& Ney. In other cases (Tempel-Tuttle, Juels-Holvorcem, and H\"{o}nig), the value of $S$ is not well-constrained. One object that seems to deviate notably and significantly from the relation that is not a JF object is Comet Wilson. The spectrum of Wilson, shown in Figures 1e and 2f of \citet{hlr94} is very unusual in that it has a shape unlike that of any other comet in their sample. 

The two YLP objects (NEAT and McNaught-Hartley) have virtually identical values of both $\Delta$m and $S$, and so their symbols overlap to the point of being indistinguishable (although both sets of error bars are visible).  And although their orbital inclinations are also nearly identical (to within 2 degrees) and might suggest a common origin, their other orbital parameters are not. 

Finally, intrinsic changes in the nature of the dust ejected with time may introduce some amount of scatter into this figure. \citet{ney74b} found that both the albedo and the strength of the silicate band in Comet Bradfield changed dramatically over a 2-week period. \citet{hlr94} noted that the band/continuum ratio of Comet Levy changed by a factor of 2 over the course of only 3 days! As described above, the associated change in $S$ for levy causes the trajectory to follow the observed $\Delta$m vs. $S$ relation, and introduces no scatter. For the data on Halley and McNaught-Hartley, however, this is not true. 

Normalizing the continuum of the higher-resolution data at 8 and 12.5 $\mu$m may have the effect of biasing the derived values of $T_{C}$ somewhat. For most silicate materials measured in the laboratory, micron-sized grains generally have a smaller value of $Q$ at 8 $\mu$m than at 12.5 $\mu$m \citep{dor95}, with the difference increasing as Mg/Fe increases. However, this would tend to bias the results toward {\it lower} values of $T_{C}$, and thus toward smaller, not larger, values of $S$. Even more interesting, the fact that the sort of Mg-rich silicates needed to match the spectral features found in comets such as Hale-Bopp \citep{dhw99}, the Giotto and Vega mass spectrometer measurements of Halley \citep{mcd91}, and in those IDPs of presumed cometary origin \citep{bhg92}, tend to have smaller emissivities at wavelengths shorter than 8 $\mu$m, including visual wavelengths, than at longer wavelengths, would make us expect that $S$ would be less than unity for all comets, which it is not.  Obviously the situation is complex and not well-understood yet. 

One possible solution is to postulate that the Fe-bearing silicates, even if they are a minority, contribute significantly to the low albedo in the visible and the infrared emissivity shortward of 8 $\mu$m. \citet{hap01,hap03} has suggested that the Fe-rich surface minerals on solar system bodies such as the Moon and asteroids have suffered weathering by the solar wind sputtering (and impact vaporization), which produce submicroscopic metallic iron. This weathering tends to darken (and redden) these materials at visible wavelengths, and wash out their band structure. Fe-bearing silicates exposed to the harsh environment of the early Sun would likely have been similarly affected. Small iron particles are found in the primitive IDPs collectively known as GEMS (Glass with Embedded Metals and Sulfides), many of which also contain silicate components that exhibit embedded cosmic ray tracks and weathered rims \citep{bhg92,brad94}, presumably left over from their exposure in interstellar space or the early solar environment. 

This is not to imply that the underlying continuum emission is dominated by small silicates, regardless of composition. Large silicate grains and carbonaceous material will exhibit a near-continuum emission, and can achieve super-equilibrium temperatures if $Q$ decreases with increasing wavelength. Small silicate grains embedded in or intimately mixed with such material will tend to be warmer than similar isolated grains. However, any isolated small Mg-rich silicate will introduce a bias toward lower a 8-13 $\mu$m color temperature than would be derived in their absence.

It is important to ask whether the deviation seen in the JF comets from the $S -\Delta$m correlation of DN objects is due primarily to their values of $S$ or to $\Delta$m. In Figure 5 and 6, we show each of these parameters plotted against $1/a$ where $a$ is the semi-major axis in AU obtained from the most recent (electronic) version of the Catalog of Cometary Orbits \citep{mar03}. On the surface it would appear that, on average, the JF comets in this sample have values of $S$ that are not that different from the mean of the rest of the sample. However, unlike the non-JF objects, which cover the entire range of $\Delta$m, none of the JF comets has a silicate band $\Delta$m in excess of 0.25 mag. An F-test of the variances of the two different populations (JF objects versus DN+YLP+OLP) indicates that they are different (probability of being the same is 0.03). A T-test of whether the mean values of the two are from the same population has a relatively high probability (0.2), but such a test is known to be unreliable if the samples have very different distributions \citep{ptvf92}. Similarly, a Kolmogorov-Smirnov test does not always yield useful results with very widely-spread distributions, and is better at detecting shifts in the median values. In this case, the JF object are in fact very close to the median of the DN+YLP+OLP sample (0.15), but are all below the mean (0.29) of the same sample due to the asymmetric nature of its distribution  (skewness of 0.7 and kurtosis of -1.2). It is not a normal distribution. The sizes of the two groups are small (4 and 14), and more data is needed to perform a deeper statistical analysis.

Nevertheless, the reason the JF objects do not fall on the $\Delta$m vs. $S$ relation in Figure 12 would seem to be mostly due to a deviation in $\Delta$m. So, while the correlation between $\Delta$m and $S$ for the majority of observed comets can be understood primarily on the basis of changes in the mean grain size, there must be an additional factor at work on the case of the JF objects. Simply put, a number of JF objects exhibit values of $S$ significantly greater than unity, while exhibiting no detectable silicate feature. The difference might be associated with a difference in mineralogy due to origin (the Kuiper-Edgewood belt for the JF objects, as opposed to the Oort cloud for the rest). \citet{aw03} have obtained long-slit spectra of the disk surrounding $\beta$ Pic, and demonstrated that in this object the silicate band is present in the inner disk, but absent further out. This would be entirely consistent with what is observed in solar system comets, given their probable source of origin. In the case of $\beta$ Pic, Weinberger et al. suggest that the small grains might be produced in the inner disk and dispersed outward by radiation pressure, but not to the most distant reaches of the disk. However, they did not detect a similar effect in the degree of crystallinity, which would argue against the grains having been close enough to the star to become annealed ($T \approx 1000 K$; \citet{sh98,sh00,hill01}).

It is also possible that these differences might be the result of weathering due to repeated passages in the inner solar system. Grains may be sorted by size during their entrainment by outflowing gas, and the largest ones may resist total removal. Due to orbital perturbations (gravitational, Poynting-Robertson drag), comets and the debris they have shed earlier can often collide. More observation of JF comets are needed.

\section{Silicate Band Structure}

Ten micron spectra at $R \sim 50-100$ now exist for about 20 comets and at least a small emission feature above a smooth blackbody is seen in most of them  (Table I). When the emission feature is strong (band/continuum $>$ 1.5), spectral structure is evident.  The individual peaks are attributed to specific crystalline and non-crystalline silicates of  olivine and pyroxene composition, and multi-component models based on measured  emissivities of crystalline and amorphous pyroxene and olivine can provide a  good match to the observed spectral shape \citep{dhw99,dhw00,hay00,har02}.

Weaker features (band/continuum $=$ 1.1-1.49), including those of many of the comets presented  here, usually have a broad trapezoidal shape lacking spectral structure. However, the lack of structure may just be the result of poorer signal-to-noise ratios (SNR) for these objects, which tend to be fainter than the majority of other objects usually studied. As described earlier, for the ISO observations of 103P/Hartley 2, \citet{col99} did not detect a silicate band, while \citet{cro99,cro00}, with slightly higher SNR, did. Our own spectrum of Hartley 2 was so poor that we did not even include it in our analysis (See Figure 7 for the data by Crovisier et al. and that obtained from the ground with BASS). Weak broad features are usually explained as arising from larger grains (grain radius of 1 $\mu$m or larger).  For example, a 2 $\mu$m or 5 $\mu$m radius disordered  (non-crystalline) olivine grain has the requisite broad shape to fit these spectral features \citep{msh87,hon04}.
 
However, even in the objects with weak silicate bands there is usually a slight peak (e.g. NEAT Fig. 1c, 3b), or at least a sharp inflection point (e.g. Kudo-Fujikawa, Fig. 3a), at 11.2 $\mu$m followed by a steep decrease from 11.2 - 12 $\mu$m consistent with the presence of crystalline olivine. For example, Comet Kudo-Fujikawa has one of the weakest silicate band features of any well-observed comet, being 5 - 8 times weaker than those with the strongest bands. Nevertheless, the spectral shape of this feature is remarkably similar to those objects with much larger bands. In Figure 8 we show the band strength after subtracting off the continuum (i.e. band/continuum minus 1). Also shown is the same parameter for Hale-Bopp (14 Oct. 1996) and Levy (12 Aug. 1990), normalized to the same overall band strength as in Kudo-Fujikawa. As is apparent in the figure, the silicate band in Kudo-Fujikawa has the same full width at half maximum and flattens out at the top at approximately the same wavelengths as the other two comets. Again, this combination of band strength and shape is consistent with the presence of a population of grains a few microns in size on top of a ÒcontinuumÓ emission. However, large silicate grains alone may not be able to explain both the silicate band emission and the continuum, as the silicate band structure and central wavelength change with increasing grain size (above 3 $\mu$m), at least for compact spheroidal particles \citep{msh87,lm00,min03}. 
 
Because a substantial fraction of the grains measured {\it in situ} for Comet Halley were organic (CHON) or an organic-silicate mix \citep{kissv86,kissg86}, the strength or weakness of the silicate band can be affected by the organic component. If the organic component has a large value of $Q_{VIS}/Q_{IR}$, such a component could be responsible for objects exhibiting large values of $S$ with little or no silicate band, that is, a ``base superheat'' which is driven even higher by small grain size (in either the organics or silicates).  More detailed modeling with a realistic particle size distribution, chemical composition,  and grain porosity will be required for a better understanding of these competing effects.
 
The spectrum of Kudo-Fujikawa, although having a silicate band as weak as those of JF objects, does not seem to be quite the same in shape. Figure 9 shows the band/continuum of this object compared to Borrelly and Hartley 2. Without any rescaling, the heights of the silicate bands in these objects are essentially identical at 11.2 $\mu$m, but the JF  objects seem to have slightly stronger emission at shorter wavelengths. More data on JF objects with greater SNR are needed before we can explore these possible differences, and the Stardust mission to Comet 81P/Wild 2 should provide us with our most detailed look at the nature of the grains in at least one JF comet.
 
Another interesting aspect of these data are their relation to the dust emission seen in the spectra of pre-main sequence stars. From surveys of the silicate band emission in there stars, it is known that while most do not exhibit spectra containing features due to crystalline silicate material (i.e. \citet{hbt98} ), a small but significant fraction do \citep{bouw01,mls99,mls04}. Figure 8 also shows the spectrum of HD 35187 (\citet{mls04}, but see also \citet{mls00}), a star often classified as a Herbig Ae star or as a Vega-like main sequence star. Its spectrum plainly shows the 10.5 and 11.2 $\mu$m peaks due to crystalline silicates, and its bandwidth and shape is very similar to those of comets Hale-Bopp and Levy. In Figure 10 we compare the spectrum of the JF comet Borrelly with that of HD 31648, an object with less distinct crystalline peak at 11.2 $\mu$m. However, the majority of the over 40 pre-main sequence objects observed by \citet{mls94,mls04} do not exhibit such comet-like spectra. In many objects, such as UX Ori, the 11.2 $\mu$m peak is ill-defined or absent, as shown in Figure 11. However, it has been suggested \citep{vda97,mls99} that only the most evolved objects (those closer to the main sequence and exhibiting inner disk clearing as evidenced by reduced dust emission at near- and mid-infrared wavelengths) exhibit significant crystallinity (required for a strong sharp 11.2 $\mu$m band), and our solar system is, after all, more evolved than the majority of PMS stars! It is therefore not too surprising that these comets resemble only a subset of the PMS population. There is independent evidence that infalling comet-like bodies exist in fairly evolved systems with silicate band emission \citep{gra97,lh88,lh89,beu90}.
 
\section{Conclusions}
We have reported previously unpublished mid-IR spectra of 5 comets: C/2002 O4 (H\"{o}nig), C/2002 X1 (NEAT), C/2002 X5 (Kudo-Fujikawa), C/2002 Y1 (Juels-Holvorcem), and 69P/Taylor. In 4 of these object (H\"{o}nig, NEAT, Kudo-Fujikawa, and Taylor) we detected the presence of the emission band due to silicates. In the other (Juels-Holvorcem) the band was weak or absent, but the quality of the data was low. Four objects (H\"{o}nig, NEAT, Taylor, and Juels-Holvorcem) contained grains radiating at temperatures generally above those expected of a blackbody at the heliocentric distance of the comet, although for any individual object the results are within 2 standard deviations of the blackbody temperature. 

In the one object where we obtained data both on and offset from the nucleus of Comet Kudo-Fujikawa, the dust emission characteristics seemed unchanged. The timing of the observations were such that they were sampling the same grains if all were traveling at a constant velocity equal to that observed in Comet Hale-Bopp. At least in this one object, any possible evidence for significant modification of the grain size by grain size sorting or disintegration was not evident. The near lack of excess temperature and weakness of the silicate band indicated that the grains ejected in Kudo-Fujikawa were not dominated by small silicate grains seen in most  ``young'' comets. However, the shape of the silicate band itself requires it be produced by whatever population of small grains that were present.

Merging this sample with objects observed earlier at similar spectral resolutions and wavelength coverage, a few notable characteristics emerge. The strong correlation between excess temperature and silicate band strength previously reported is confirmed here for the majority of the dynamically new objects in our sample. However, there is a tendency for Jupiter family objects to deviate from this correlation in the sense that they tend to have temperature characteristics close to the middle of the range exhibited by the sample as a whole, but silicate bands that are consistently weaker than the average. If this trend is confirmed with a larger sample, then it would suggest that our solar system possessed a radial gradient in the size distribution of its silicate grains, similar to that observed in $\beta$ Pic today.

It is apparent that the spectra of solar system comets possess many features in common. Most well-observed objects exhibit temperatures in excess of that expected from a blackbody at the equilibrium temperature expected for their heliocentric distances. Most show a silicate emission feature, although the strength can be quite different from object to object, or even from epoch to epoch for a single object. When strong enough to measure reliably, the silicate band usually has a trapezoidal shape above the continuum, usually with a rather distinct peak at 11.2 $\mu$m feature attributed to crystalline olivine. However, it would be a mistake to say that they are all alike (Figure 12).

We also find that in some cases, most notably that of Comet Kudo-Fujikawa, even though its silicate band is weak, the shape and structure after subtracting the continuum is similar to that of other dynamically new and long period comets. It also resembles the emission of more evolved pre-main sequence stars, such as HD 35187. For the same band strength at 11.2 $\mu$m, the 9-10 $\mu$m fluxes  of the JF comets tend to be slightly larger than that of Kudo-Fujikawa (and other dynamically new and long period comets), and are more similar to pre-main sequence stars that are less evolved than HD 35187. It would seem that the dynamically new and long period comets, which come from the Oort Cloud but whose origins were probably inside the orbit of Neptune, were endowed with grains of a somewhat different nature than the Jupiter family objects, which probably originated from the Kuiper-Edgeworth Belt beyond the orbit of Neptune. During the first tens of millions of years of the evolution of the solar system, these two regions of comet formation would have been processed over time to various degrees with solar wind exposure, silicate grain annealing, small grain expulsion, grain coagulation, and settling of grains in the protosolar disk. Alien observers (if they existed) might, at various times, have seen a system that progressively evolved from one that looked like UX Ori to HD 31648 to HD 35187.

\acknowledgments

For this work MLS was supported in part by NASA grant NAG5-9475 and the University Research Council of the University of Cincinnati. DKL and RWR were supported by The Aerospace Corporation's Independent Research and Development program and by the United States Air Force Space and Missile Systems Center through the Mission Oriented Investigation and Experimentation program, under contract F4701-00-C-0009. The authors wish to thank the IRTF Director, Alan Tokunaga, for allowing them to carry out the daytime observations of Comet Kudo-Fujikawa, Bill Golisch for the extra hours he spent as the telescope operator on those lengthy observing sessions, and the anonymous referee for their many insightful comments.

\clearpage

 \begin{figure}
 \center
\plotone{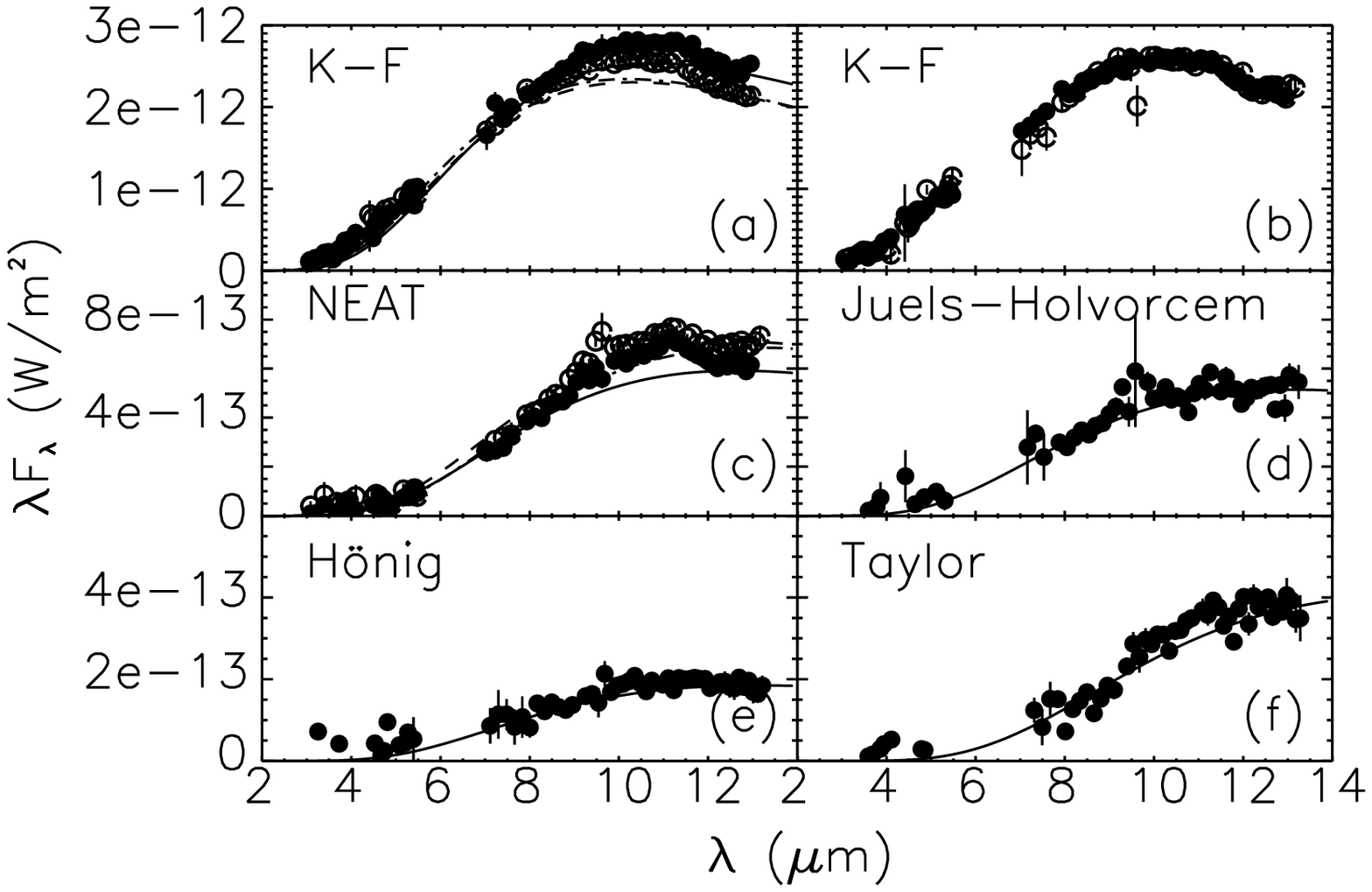}
\caption{(a) - The observed flux of Comet Kudo-Fujikawa for January 9 (filled circles) and 10 (open circles), 2003. The lines refer to the graybody curve, normalized to the flux at 7.5-8.5 $\mu$m and 12-13 $\mu$m. Solid line: January 9 data with T=340 K; dashed line: January 10 data with T=355 K; dot-dashed line:  January 10 data with T=360 K.  (b) The flux of Comet Kudo-Fujikawa from January 10, centered on the nucleus (filled circles) and in the coma (open circles) 2.6 arcsec off the nucleus (the aperture radius was 1.7 arcsec). The flux for the coma has been scaled upward by a factor of 2.6 . (c) The spectrum of Comet NEAT on January 9 (filled circles) and January 10 (open circles), 2003. The graybody curve for the January 9 data has T=290 K (solid line). For the January 10 observations, the  T=280 K (dashed line) and T=290 K (barely visible dot-dashed line) graybody curves are both shown. (d) Comet Juels-Holvorcem and the graybody continuum with T-290 K (solid line). (e) Comet H\"{o}nig from August 1, 2002, along with a graybody continuum with T=280 K (solid line). (f) Comet Taylor on February 9, 1998, and the continuum with T=230 K.\label{fig1}}
\end{figure}
 
\clearpage

\clearpage

 \begin{figure}
\plotone{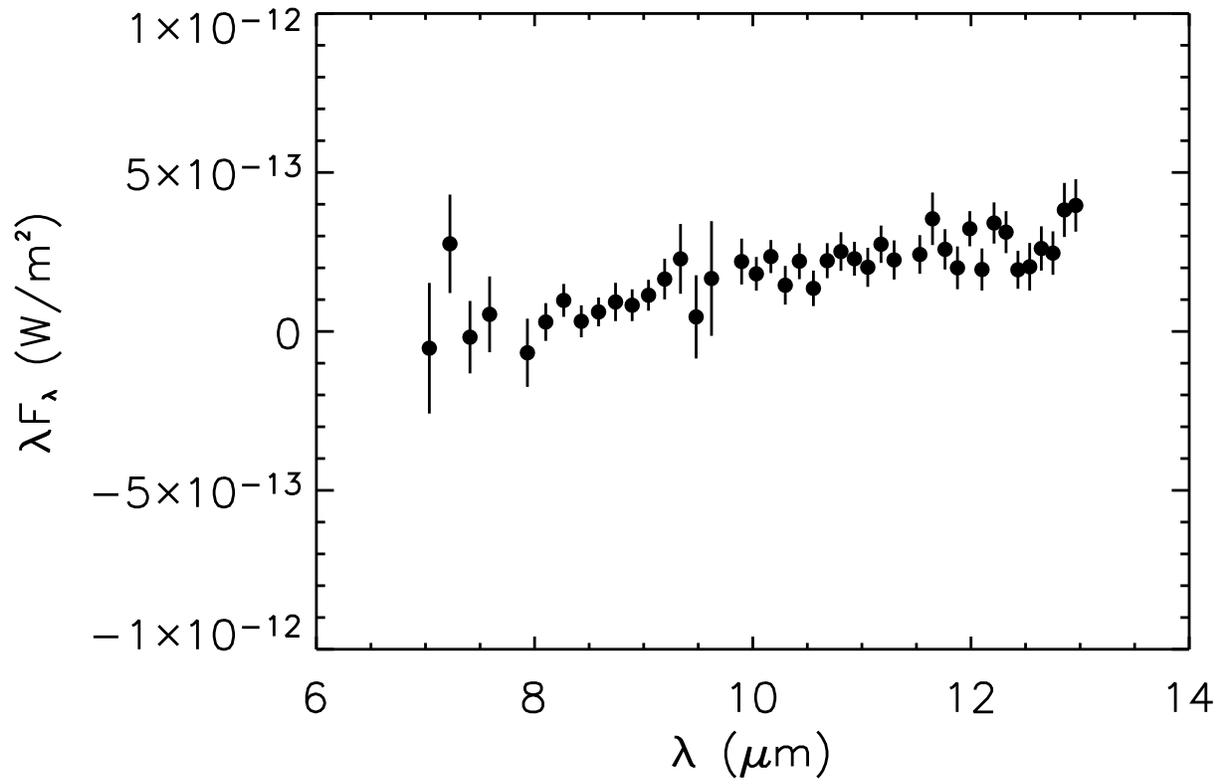}
\caption{The difference in flux for Comet Kudo-Fujikawa between January 9 and 10. \label{fig2}}
\end{figure}
 
\clearpage

 \begin{figure}
\plotone{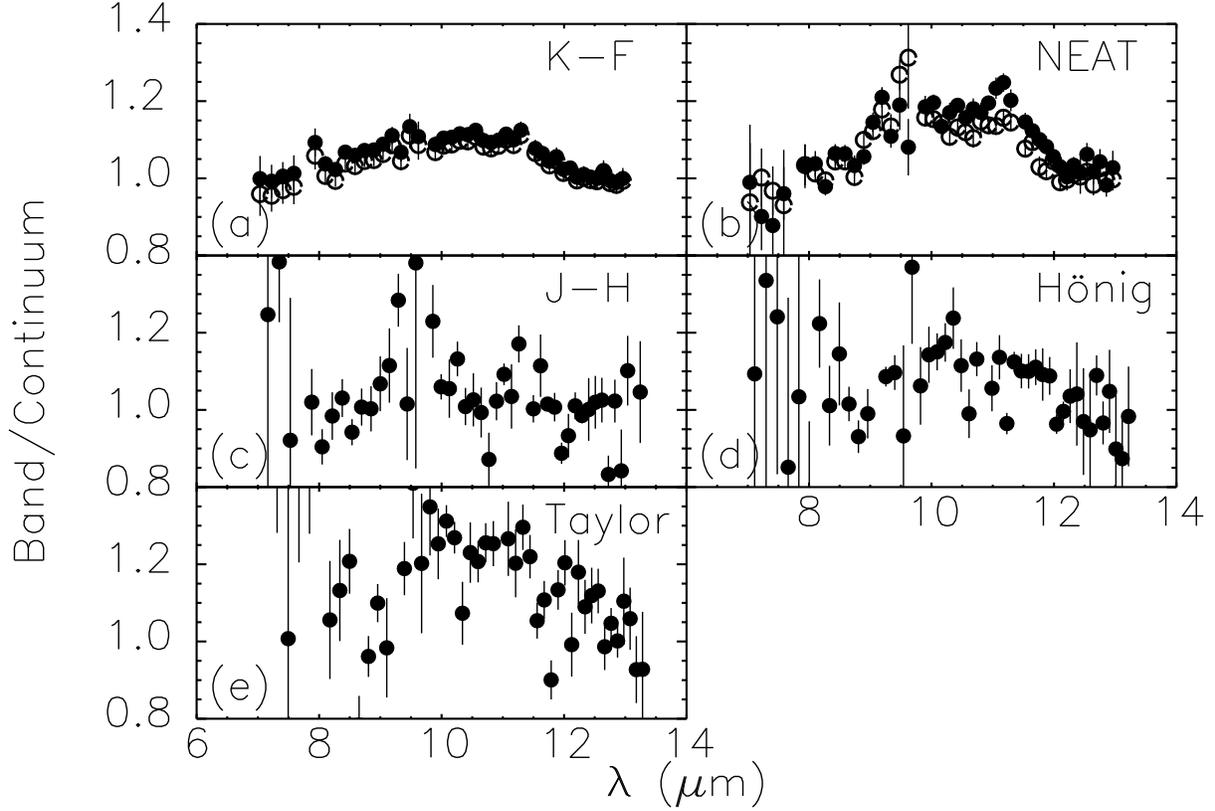}
\caption{(a) - The ratio of the observed flux in the center of the silicate band, divided by the continuum flux for Comet Kudo-Fujikawa on January 10 (the curve is almost identical for the previous night). The filled circles show the resulting values assuming $T_{C} = 355$ K, while the open circles were derived using $T_{C} = 360$.  (b) - The band/continuum ratio for Comet NEAT on January 9 and 10 (filled and open circles, respectively). (c), (d), and (e) - The band/continuum ratio for Comet Juels-Holvorcem, Comet H\"{o}nig, and Comet Taylor, respectively. \label{fig3}}
\end{figure}

\clearpage

\begin{figure}
\plotone{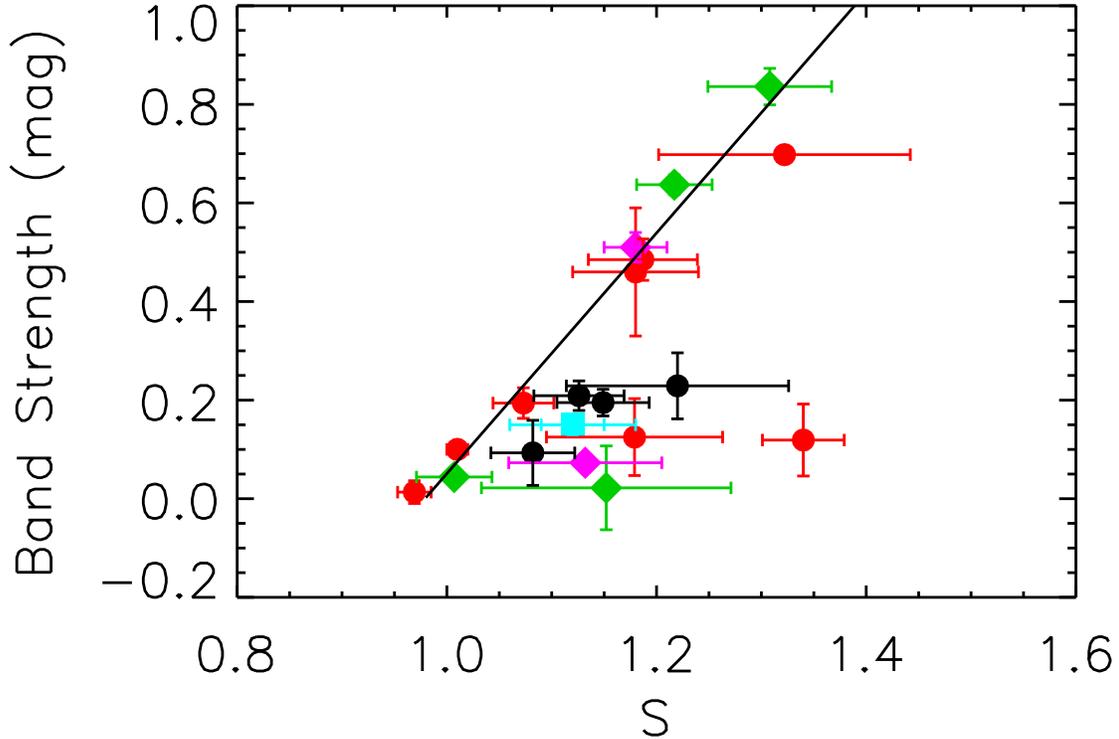}
\caption{The silicate band strength, expressed as a magnitude difference between the in-band flux and continuum flux interpolated to the same wavelength regime, plotted versus the ratio of the continuum temperature to that of a blackbody at the equilibrium temperature expected for the heliocentric distance of the observation (ÒsuperheatÓ). Red circles: dynamically new objects; cyan squares: young long-period comets; green diamonds: old long-period comets; magenta diamonds: Halley family comets; black circles: Jupiter family comets.  With the exception of H\"{o}nig, which has large error bars, and Wilson, whose spectrum is unusual (see text), all of the dynamically new comets define a tight relation between silicate band strength and excess temperature, as discussed by \citet{gn92}. The straight line is the best fit to the data in our sample, excluding the Jupiter family objects and the peculiar outlier Comet Wilson (see discussion). All of the Jupiter family objects fall to the right of this curve. Two other Jupiter family objects (Tempel 1 and Tempel 2) must also inhabit this region of parameter-space, as both exhibit a temperature excess, but no discernable silicate feature \citep{dkl95}. \label{fig4}}
\end{figure}
 
\clearpage

\begin{figure}
\plotone{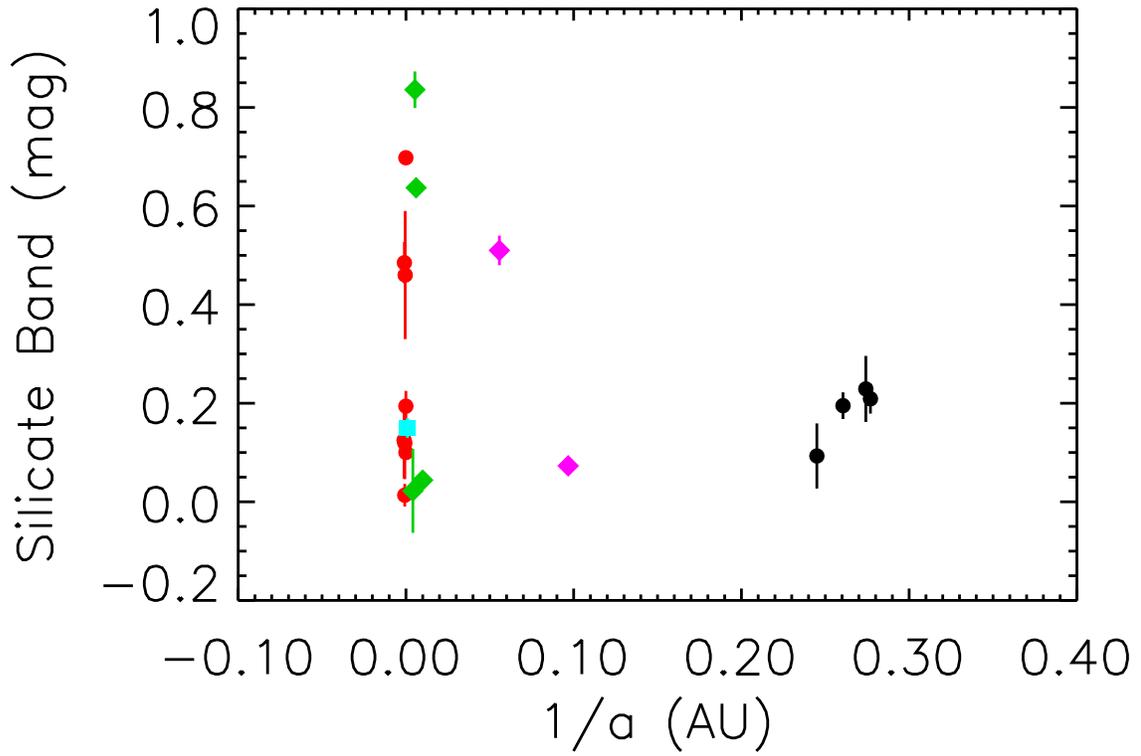}
\caption{The silicate band strength versus 1/a (AU$^{-1}$) for this sample. The colors and symbols are the same as in Figure 4. The Jupiter family objects are restricted to the small-band region of the figure. \label{fig5}}
\end{figure}
 
\clearpage

\begin{figure}
\plotone{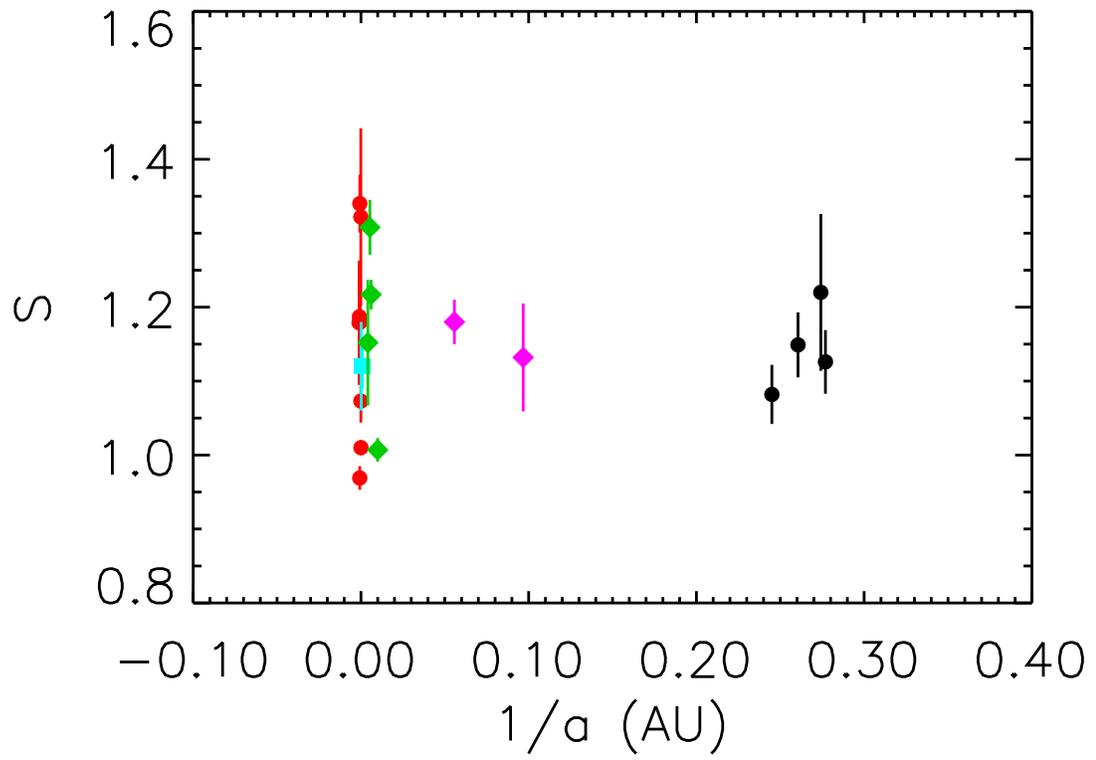}
\caption{The excess temperature versus 1/a, with the same color and symbols as previous figures. Objects from all orbital classes exhibit a range of values in excess temperature. \label{fig6}}
\end{figure}
 
\clearpage

\begin{figure}
\plotone{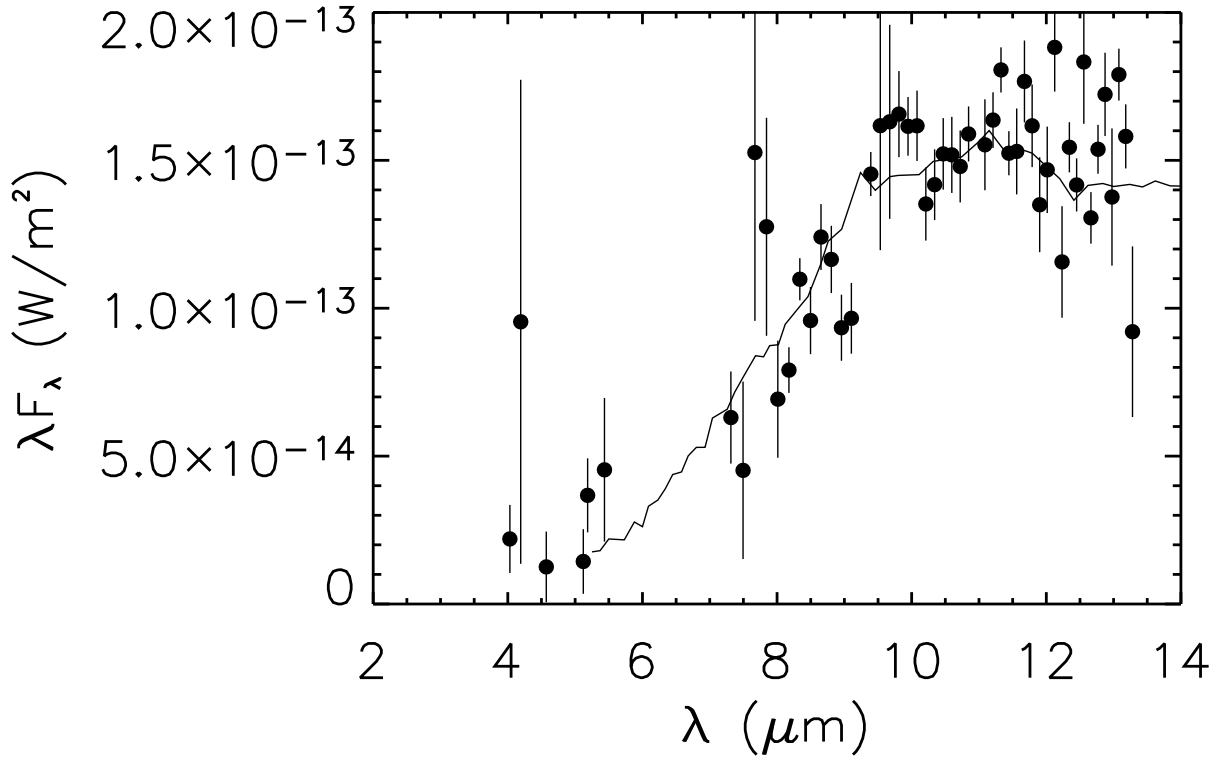}
\caption{The spectrum of 103P/Hartley 2 obtained with BASS on 9 Feb. 1998 and a heliocentric distance of 1.2 AU (filled circles) and that obtained with ISOCAM-CVF on 31 Dec. 1997 \citep{cro00} at 1.0 AU (solid line). The ISO data (extracted from the camera images) has been scaled down by a factor of 14 to that obtained with the small entrance aperture of BASS. \label{fig7}}
\end{figure}
 
\clearpage

\begin{figure}
\plotone{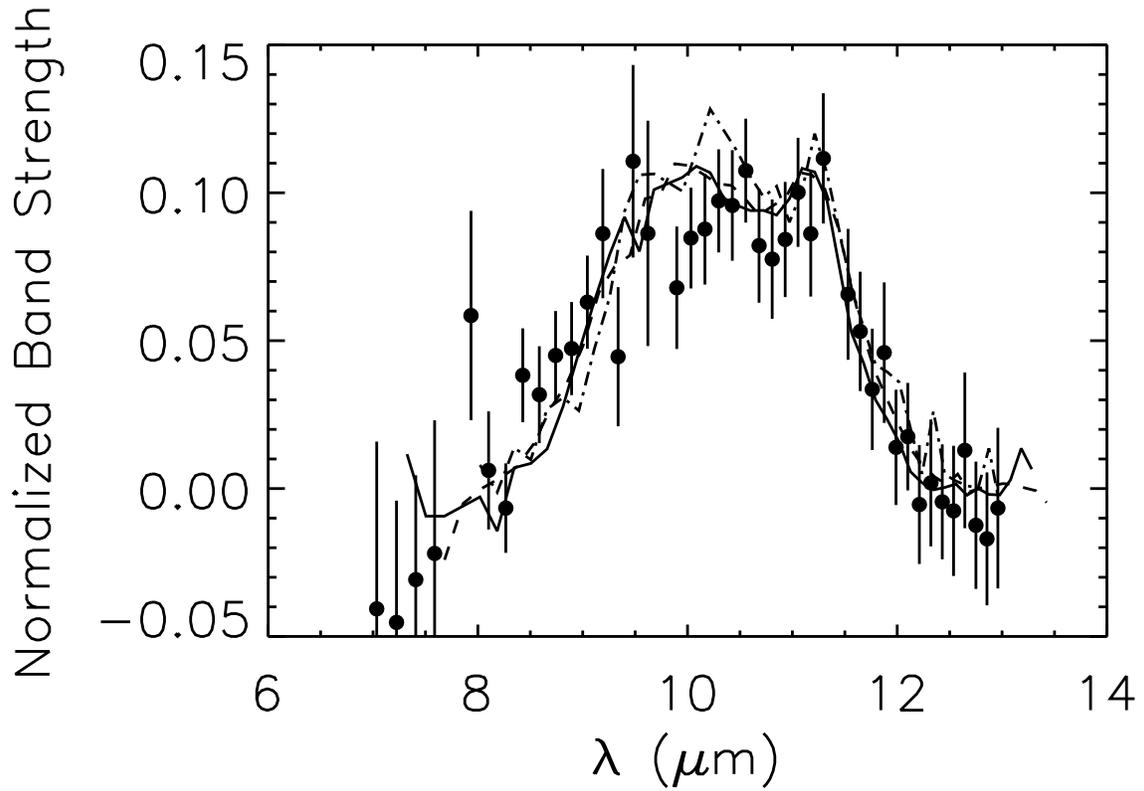}
\caption{The continuum-subtracted silicate feature of Comet Kudo-Fujikawa (filled circles), compared to those of Hale-Bopp (solid line), Levy (dashed line) and the star HD 35187 (dot-dashed line). \label{fig8}}
\end{figure}
 
\clearpage

\begin{figure}
\plotone{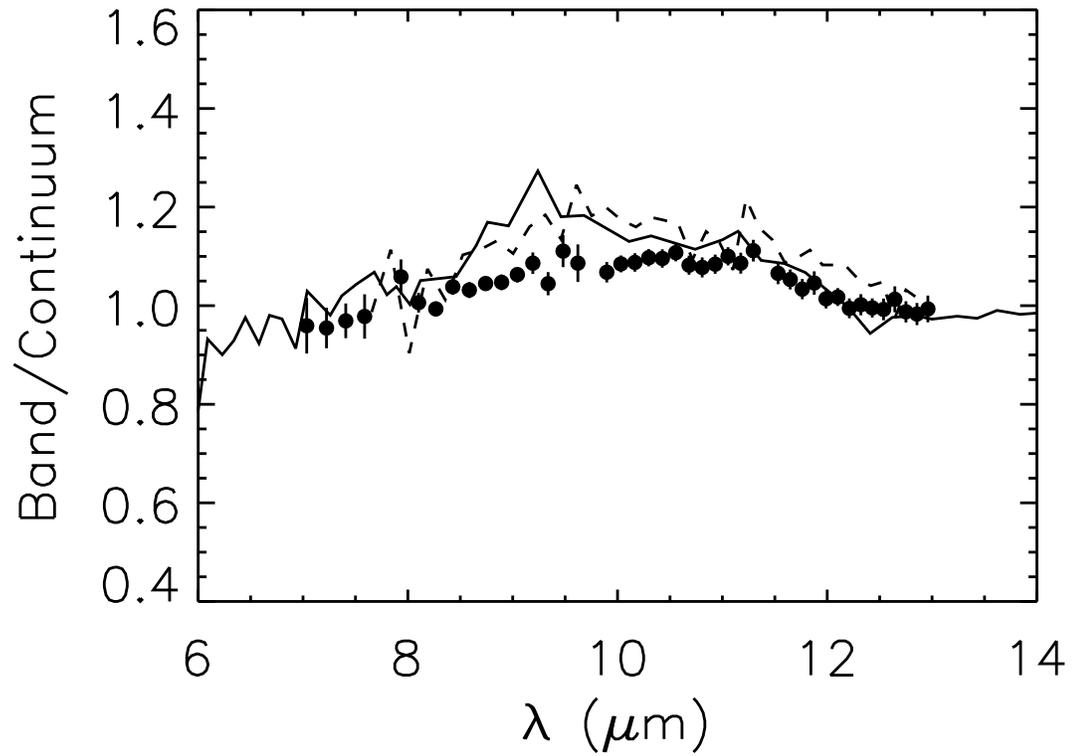}
\caption{The band/continuum ratio for Hartley 2 (solid line) Borrelly (dashed line) and Kudo-Fujikawa. The data for Hartley 2 and Borrelly were taken from \citet{cro00} and \citet{msh96}, respectively. \label{fig9}}
\end{figure}
 
\clearpage

\begin{figure}
\plotone{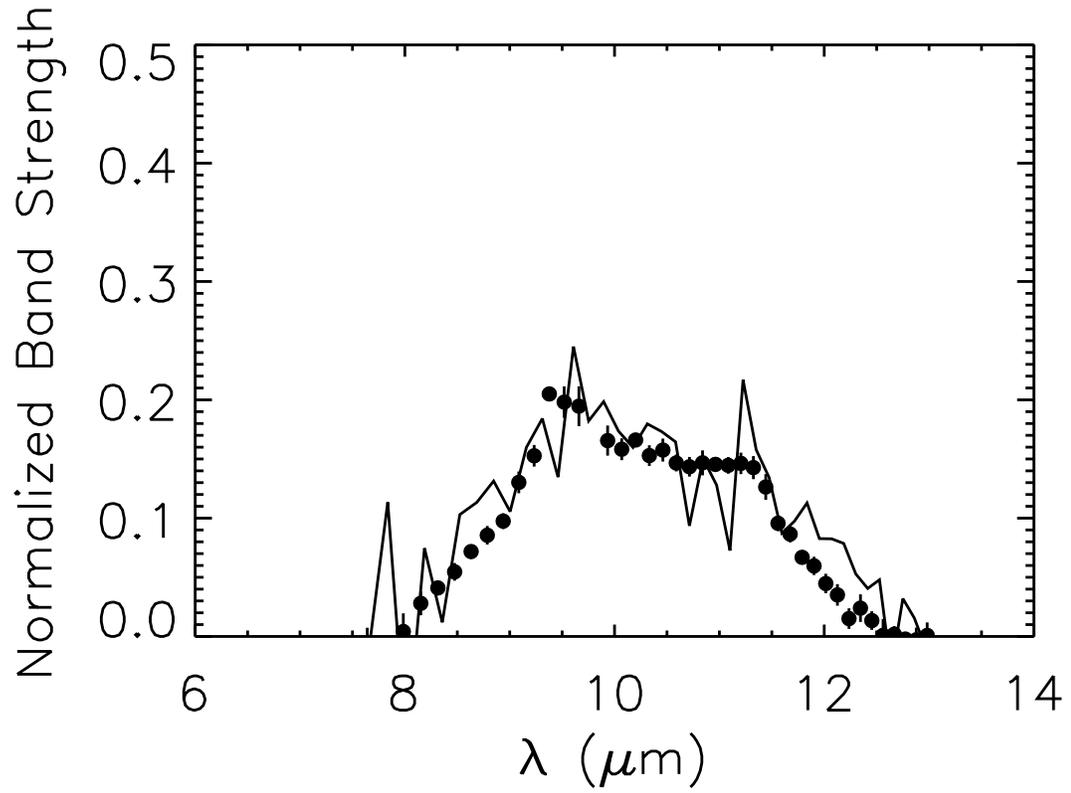}
\caption{The continuum-subtracted (band/continuum $-$ 1) for Borrelly (solid line) and HD 31648 (filled circles), scaled to the same height at 10.5 $\mu$m. The spectrum of HD 31648 is from \citet{mls99,mls04}. \label{fig10}}
\end{figure}
 
\clearpage

\begin{figure}
\plotone{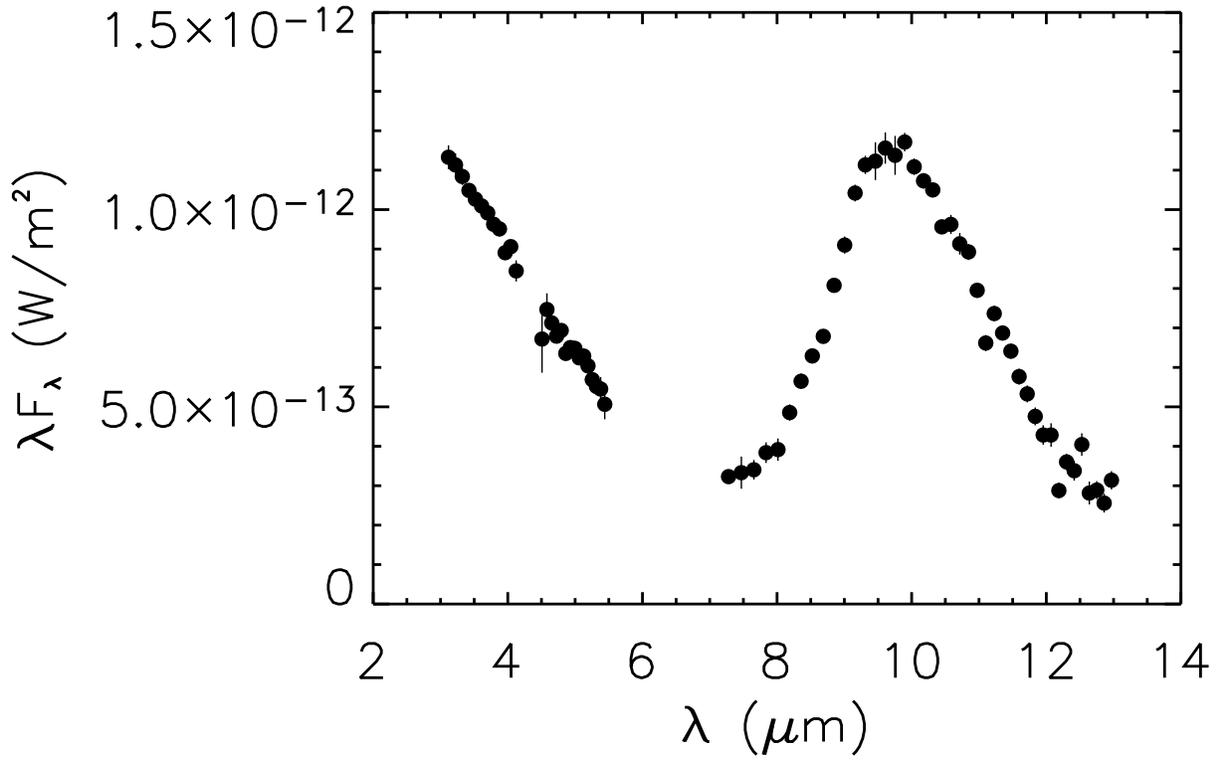}
\caption{The BASS spectrum of the HAEBE stars UX Ori, obtained on 13 Dec. 1994 (from \citet{mls04}). \label{fig11}}
\end{figure}
 
\clearpage

\begin{figure}
\plotone{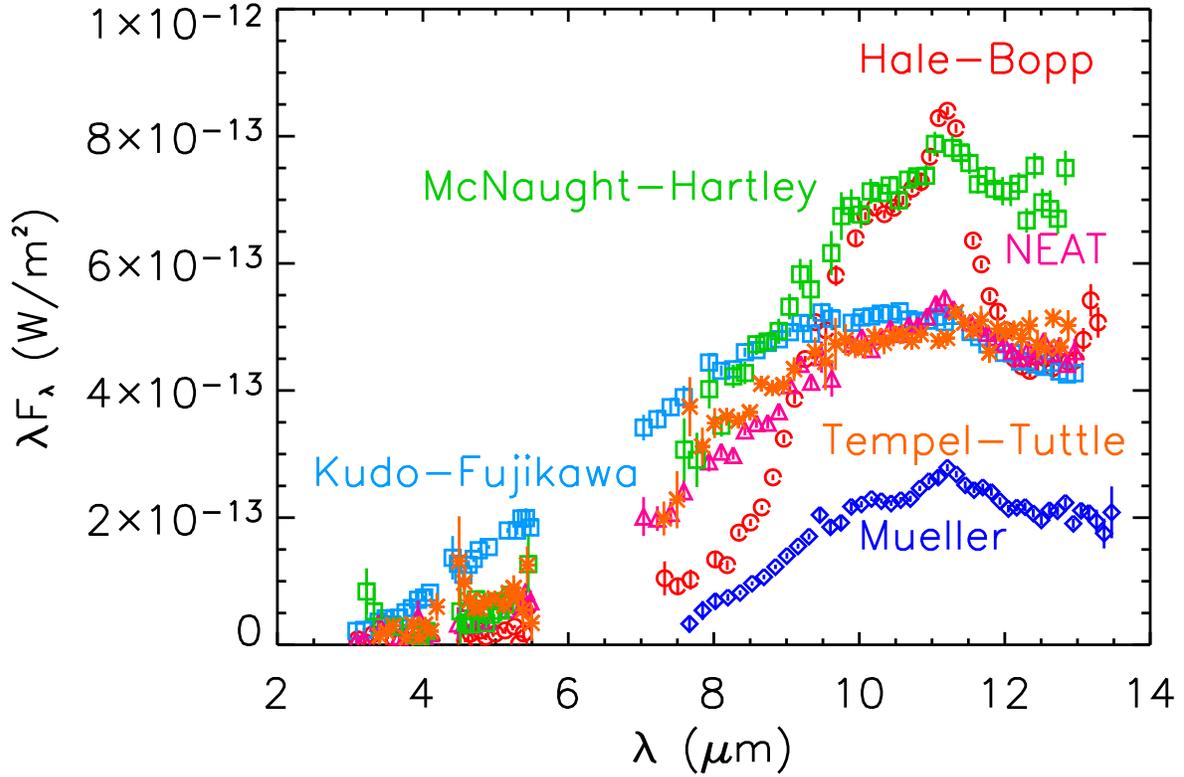}
\caption{Emission spectra of comets. The flux levels of Hale-Bopp and Kudo-Fujikawa have been scaled downward by a factor of in order to fit within the plot. Data are from \citet{msh97}[Hale-Bopp], \citet{hlr94}[Mueller], \citet{dkl00}[Tempel-Tuttle], \citet{dkl02}[McNaught-Hartley], and this paper [Kudo-Fujikawa \& C/2002 V1 NEAT]. \label{fig12}}
\end{figure}

\clearpage

\begin{deluxetable}{lrccccc}
\tabletypesize{\scriptsize}
\tablecaption{New Comet Observations \label{tbl-1}}
\tablewidth{0pt}
\tablehead{
\colhead{Comet} & 
\colhead{UT Date}  &
\colhead{UT Time} &
\colhead{Airmass Range} &
\colhead{Mean Airmass} &
\colhead{Calibrator} &
\colhead{Calibrator Aimass} }  

\startdata
Taylor (69P) & 9 Feb. 1998 & 0901-0992 & 1.03-1.04 & 1.03 & $\alpha$ Lyr & 1.25 \\
H\"{o}nig (C/2002 O4) & 1 Aug. 2002 & 1246-1319 & 1.16 & 1.16 & $\alpha$ Lyr  & 1.22 \\
NEAT (C/2002 V1) & 9 Jan. 2003 & 0422-0559 & 1.03-1.21 & 1.10 & $\alpha$ Tau & 1.04 \\
NEAT (C/2002 V1) & 10 Jan. 2003 & 0608-0724 & 1.31-1.79 & 1.55 & $\alpha$ CMa & 1.54 \\ 
Kudo-Fujikawa (C/2002 X5) &  9 Jan. 2003 & 0219-0312 & 1.94-2.85 & 2.32 & $\alpha$ CMa & 2.39 \\
Kudo-Fujikawa (C/2002 X5) &  10 Jan. 2003 & 0120-0256 & 1.42-2.21 & 1.79 & $\alpha$ Tau & 1.71 \\
Juels-Holvorcem (C/2002 Y1) & 20 Feb. 2003 & 1415-1536 & 2.24-1.95 & 2.97 & $\alpha$ Lyr & 2.25 \\
 \enddata

\end{deluxetable}

\begin{deluxetable}{llrrrrrrrr}
\rotate
\tabletypesize{\scriptsize}
\tablecaption{Comet Parameters \label{tbl-2}}
\tablewidth{0pt}
\tablehead{
\colhead{Comet} & \colhead{1/a (AU$^{-1}$)}   & \colhead{Class\tablenotemark{a}}  &
\colhead{Date\tablenotemark{b}} &
\colhead{$r$(AU)} & \colhead{$T_{C}$\tablenotemark{c}} & \colhead{$T_{BB}$\tablenotemark{d}} &
\colhead{$S$\tablenotemark{e}} & \colhead{$Band/Cont$\tablenotemark{f}}  &
\colhead{$\Delta$m B/C\tablenotemark{g}}}   

\startdata
 & & & \it{Newly Observed} & & & \\
H\"{o}nig (C/2002 O4) & -0.001102 & DN & 1.5 Aug. 2002 & 1.367 & 280$\pm$20 & 237 & 1.18$\pm$0.08 & 1.12$\pm$0.08 & 0.12$\pm$0.08 \\
Kudo-Fujikawa (C/2002 X5) & 0.0 & DN & 9.1 Jan. 2003 & 0.67 & 340$\pm$5 & 340 & 1.00$\pm$0.01 & 1.09$\pm$0.01 & 0.09$\pm$0.01\\
     &    &    & 10.1 Jan. 2003 & 0.63 & 355$\pm$5 & 350 & 1.02$\pm$0.01 & 1.11$\pm$0.01 & 0.11$\pm$0.01 \\
 NEAT (C/2002 V1) & 0.000876 & YLP & 9.2 Jan. 2003 & 1.197 & 290$\pm$10 & 254 & 1.14$\pm$0.04 & 1.17$\pm$0.02 & 0.17$\pm$0.02 \\
    &   &   &   10.3 Jan. 2003 & 1.174 & 280$\pm$10 & 257 & 1.09$\pm$0.04 & 1.13$\pm$0.02 & 0.13$\pm$0.02 \\
    Juels-Holvorcem (C/2002 T1) & 0.004135 & OLP & 20.6 Feb. 2003 & 1.219 & 290$\pm$30 & 252 & 1.15$\pm$0.12 & 1.02$\pm$0.08 & 0.02$\pm$0.08 \\
 Taylor (69P) & 0.27416 & JF & 9.4 Feb. 1998 & 2.174 & 230$\pm$20 & 189 & 1.22$\pm$0.11 & 1.23$\pm$0.08 & 0.22$\pm$0.07 \\
  & & & \it{Previously Reported} & & & \\
  Hale-Bopp (C/1995 O1) & 0.005359 & OLP & 14 Oct. 1996 & 2.73 & 220$\pm$10 & 168 & 1.31$\pm$0.06 & 2.16$\pm$0.07 & 0.83$\pm$0.04 \\
  Kohoutek (C/1973 E1) & -0.000056  &  DN & 21 Dec. 1973 & 0.31 & 660$\pm$60 & 499 & 1.32$\pm$0.12 & 1.89$\pm$0.01 & 0.69$\pm$0.01 \\
  Halley (1P) & 0.055736 & HF & 17 Dec. 1985 & 1.25 & 300$\pm$10 & 249 & 1.21$\pm$0.04 & 1.64$\pm$0.04 & 0.53$\pm$0.03 \\
    &  &  & 16 Jan. 1986 & 0.79 & 360$\pm$5 & 313 & 1.15$\pm$0.02 & 1.58$\pm$0.03 & 0.50$\pm$0.02\\
    Bradfield (C/1987 P1) & 0.006058 & OLP & 16 Dec. 1985 & 0.99 & 340$\pm$10 & 279 & 1.22$\pm$0.04 & 1.80$\pm$0.03 & 0.64$\pm$0.02 \\
    Mueller (C/1993 A1) & -0.000918 & DN & 16 Nov. 1993 & 2.06 & 230$\pm$10 & 194 & 1.19$\pm$0.05 & 1.56$\pm$0.06 & 0.48$\pm$0.04 \\
    Levy (C/1990 K1) & -0.000444 & DN & 12 Aug. 1990 & 1.56 & 270$\pm$10 & 223 & 1.22$\pm$0.04 & 1.72$\pm$0.04 & 0.59$\pm$0.03 \\
     & & & 15 Aug. 1990 & 1.51 & 255$\pm$5 & 226 & 1.13$\pm$0.02 & 1.37$\pm$0.06 & 0.34$\pm$0.05 \\
     Wilson (C/1986 P1) & -0.000664 & DN & 24 April 1987 & 1.20 & 340$\pm$10 & 254 & 1.34$\pm$0.04 & 1.12$\pm$0.08 & 0.12$\pm$0.07 \\
     Austin (C/1989 X1) & -0.000700 & DN & 6 May 1990 & 0.78 & 305$\pm$5 & 315 & 0.97 $\pm$0.02 & 1.13$\pm$0.02 & 0.13$\pm$0.02 \\
     McNaught-Hartley (C/1999 T1) & 0.000173 & YLP & 31 Jan. 2001 & 1.40 & 250$\pm$10 & 235 & 1.06$\pm$0.04 & 1.18$\pm$0.05 & 0.18$\pm$0.05 \\
      & & & 1 Feb. 2001 & 1.41 & 275$\pm$5 & 234 & 1.17$\pm$0.02 & 1.15$\pm$0.02 & 0.15$\pm$0.02 \\
      Okazaki-Levy-Rudenko (C/1989 Q1) & -0.000031 & DN & 7 Nov. 1989 & 0.65 & 370$\pm$10 & 345 & 1.07$\pm$0.03 & 1.20$\pm$0.03 & 0.19$\pm$0.03 \\
      Faye (4P) & 0.260557 & JF & 17 Nov. 1991 & 1.51 & 260$\pm$10 & 226 & 1.15$\pm$0.04 & 1.19$\pm$0.03 & 0.19$\pm$0.03 \\
      Borrelly (19P) & 0.276914 & JF & 13 Dec. 1994 & 1.45 &260$\pm$10 & 231 & 1.13$\pm$0.04 & 1.21$\pm$0.03 & 0.21$\pm$0.03 \\
      IRAS-Araki-Alcock (C/1983 H1) & 0.009971 & OLP & 11 May 1983 & 1.00 & 280$\pm$10 & 278 & 1.01$\pm$0.04 & 1.04$\pm$0.02 & 0.04$\pm$0.02 \\
      Tempelk-Tuttle (55P) & 0.096735 & HF & 8 Feb. 1998 & 1.03 & 310$\pm$20 & 274 & 1.13$\pm$0.07 & 1.07$\pm$0.02 & 0.07$\pm$0.02 \\
      Schaumasse (24P) & 0.244992 & JF & 7 Feb. 1993 & 1.24 & 270$\pm$10 & 250 & 1.08$\pm$0.04 & 1.09$\pm$0.07 & 0.09$\pm$0.07 \\
 \enddata

\tablenotetext{a}{DN: $1/a < 50x10^{-6}$ AU$^{-1}$ 
YLP: $50x10^{-6} < 1/a < 2000x10^{-6}$ AU$^{-1}$ 
OLP: $2000x10^{-6} < 1/a < 2.9x10^{-2}$ AU$^{-1}$ (P$ > 200$ yr)
HF: P$< 200$ yr and T$_{J} < 2$ 
JF: P$< 200$ yr and T$_{J} >2$}
\tablenotetext{b} {For previously observed objects, the dates given may refer to the mean date of more than one epoch, where the data have been merged.}
\tablenotetext{c} {$T_{C}$ is defined as the temperature of the underlying continuum (assumed to be planckian) determined for the spectral regions from 7.5-8.5 $\mu$m and 12-13 $\mu$m.}
\tablenotetext{d} {$T_{BB}$ is the temperature a true blackbody would have in equilibrium at the heliocentric distance at the time of observation.}
\tablenotetext{e}{ S is the ratio $T_{C}/T_{BB}$.}
\tablenotetext{f} {Band/Cont is the ratio of the flux between 10.0-11.0 $\mu$m to that of the underlying continuum.}
\tablenotetext{g} {$\Delta$m (B/C) is the Band/Cont expressed as a difference in magnitude.}

\end{deluxetable}

\end{document}